\documentclass[journal, draftcls, one column, 12pt]{IEEEtran}
\usepackage{amsmath}
\usepackage{graphicx}
\usepackage{amssymb}
\usepackage{cases}
\usepackage{cite}
\usepackage[ruled,vlined,lined,ruled,linesnumbered]{algorithm2e}
\usepackage{relsize}
\usepackage{multirow}
\usepackage[nodisplayskipstretch]{setspace}
\usepackage[top=1in, bottom=1in, left=0.7in, right=0.7in]{geometry}
\usepackage[usenames, dvipsnames]{color}
\usepackage{epstopdf}

\renewcommand{\overrightarrow}{\boldsymbol}

\begin{document}
\title{Graph-based Cyber Security Analysis of State Estimation in Smart Power Grid}
\author{Suzhi~Bi and Ying Jun (Angela) Zhang
\thanks{S.~Bi is with the  College of Information Engineering, Shenzhen University, Shenzhen, China (E-mail: bsz@szu.edu.cn).}
        \thanks{Y.~J.~Zhang is with the Department of Information Engineering, The Chinese University of Hong Kong, Shatin, New Territories, Hong Kong, and Shenzhen Research Institute, The Chinese University of Hong Kong, Shenzhen, China. (Email: yjzhang@ie.cuhk.edu.hk).}}
\maketitle

\vspace{-1.8cm}

\section*{Abstract}
Smart power grid enables intelligent automation at all levels of power system operation, from electricity generation at power plants to power usage at households. The key enabling factor of an efficient smart grid is its built-in information and communication technology (ICT) that monitors the real-time system operating state and makes control decisions accordingly. As an important building block of the ICT system, power system state estimation is of critical importance to maintain normal operation of the smart grid, which, however, is under mounting threat from potential cyber attacks. In this article, we introduce a graph-based framework for performing cyber-security analysis in power system state estimation. Compared to conventional arithmetic-based security analysis, the graphical characterization of state estimation security provides intuitive visualization of some complex problem structures and enables efficient graphical solution algorithms, which are useful for both defending and attacking the ICT system of smart grid. We also highlight several promising future research directions on graph-based security analysis and its applications in smart power grid.

\section{Introduction}
\begin{figure}
\centering
  \begin{center}
    \includegraphics[width=0.7\textwidth]{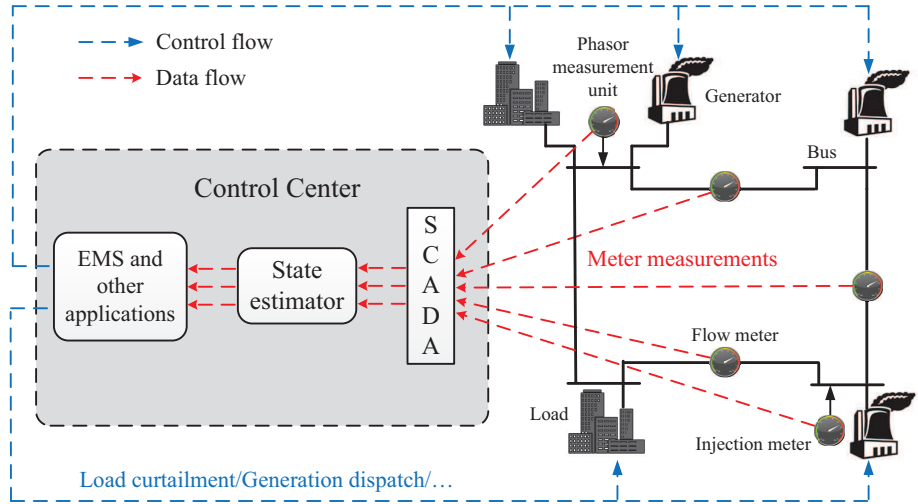}
  \end{center}
  \caption{An illustration of the operation of the SCADA/EMS system for a four-bus network. }
  \label{75}
\end{figure}

Smart power grid is committed to providing stable, high-quality and inexpensive electricity supply to meet the surging power demand of modern society through its intelligent energy management in power generation, transportation and distribution, and its introduced competitive market mechanisms. Essentially, the intelligence of smart grid is driven by its embedded ICT infrastructure, especially the EMS/SCADA (Energy Management System and Supervisory Control and Data Acquisition) system \cite{2004:Abur}. As shown in Fig.~\ref{75}, the SCADA system is responsible for collecting the measurement data reported by distributed meters/sensors, which is then fed to the state estimator located at the control center for deriving the estimation of system state variables, e.g., bus voltage amplitudes and phases. Based on the estimation, the EMS, as well as other power system applications, then makes control decisions, e.g., optimal power flow, load curtailment, and electricity pricing, to adjust the physical aspects of the power grid. Evidently, a secure and efficient power system requires accurate state estimation that truthfully reflects the system operating state.

The dependence of smart grid on its ICT infrastructure makes cyber-attacks on state estimation a viable approach to impact the normal system operation. In the conventional power network, power devices are isolated from the public network and under close control by the industrial system operator. In smart grid, however, many distributed smart meters are installed in the households, which often connect to the public internet and run IP-based communication protocols to facilitate two-way information exchange between the users and system operator. This computer-network-alike ICT structure achieves low management cost, but also exposes the smart grid to potential cyber attacks through the public information access points. One common cyber attack in smart grids is false-data injection, which distorts the measurements collected by the system operator through either physical device compromise or remote cyber-data injection \cite{2009:Liu}. Being able to compromise the state estimation, an adversary capable of false-data injection can have large impact on the power system and beyond, such as earning lucrative profit from electricity price manipulation in the power market \cite{2013:Bi,2011:Kosut}, or causing regional blackout to induce chaos and financial loss \cite{2011:Yuan}.

State estimator commonly uses bad data detection (BDD) mechanism to filter faulty data, either caused by random network error or malicious injection \cite{2004:Abur}. However, BDD is unable to detect some structured collaborating injection attacks that are disguised as normal measurements \cite{2009:Liu}. One countermeasure is data-driven detection, which uses the statistical features of the previously collected measurement data to identify anomalous measurements \cite{2011:Kosut}. Nonetheless, it cannot fully eliminate the threat of injection attacks and its performance highly depends on the accuracy of the extracted statistical features. To fundamentally mitigate false-data injection attack, it is necessary to secure meter measurements themselves to evade malicious injections by, for example, guards, video monitoring, or tamper-proof communication systems \cite{2013:Huang}. In a large power network with hundreds of meter measurements, it is tempting to devise a \emph{strategic} protection that achieves system security requirement with low cost, e.g., small number of secured devices.

\emph{Arithmetic} and \emph{graphical} methods are two popular approaches for security analysis in power system state estimation. Specifically, arithmetic approach applies algebra and matrix theory to analyze the solution space of the state estimation, and thus the potential threats and countermeasures of injection attacks (e.g., \cite{2010:Bobba,2011:Bi}). Despite its effectiveness in extensive applications, arithmetic approach is found inefficient in handling some complex problems especially for those with combinatorial features, e.g., involving selecting $k$ out of $K$ buses. Alternatively, graph-based approach, which uses graph models to characterize the security problems, can provide intuitive visualization of complex problem structures (e.g., \cite{2014:Bi,2012:Sou,2012:Rahman,2014:Bi1}). Its useful insight can lead to efficient optimal or sub-optimal graphical solution algorithms that are otherwise not achievable by arithmetic approaches. However, classic graph algorithms often need significant modifications to solve power system security problems of unique graphical structures.

In this article, we provide an overview of graphical methods for performing cyber-security analysis in power system state estimation. Specifically, we first describe the method to model power network in a graph. Then, we establish a graph-based characterization of state estimation security, and introduce some representative graphical algorithms to solve security problems in state estimation. We also suggest several future research directions on graph-based security analysis and its applications in smart power grid. Finally, we conclude this article.

\section{Graph Modeling of Power Network and Measurements}
\begin{figure}
\centering
  \begin{center}
    \includegraphics[width=0.8\textwidth]{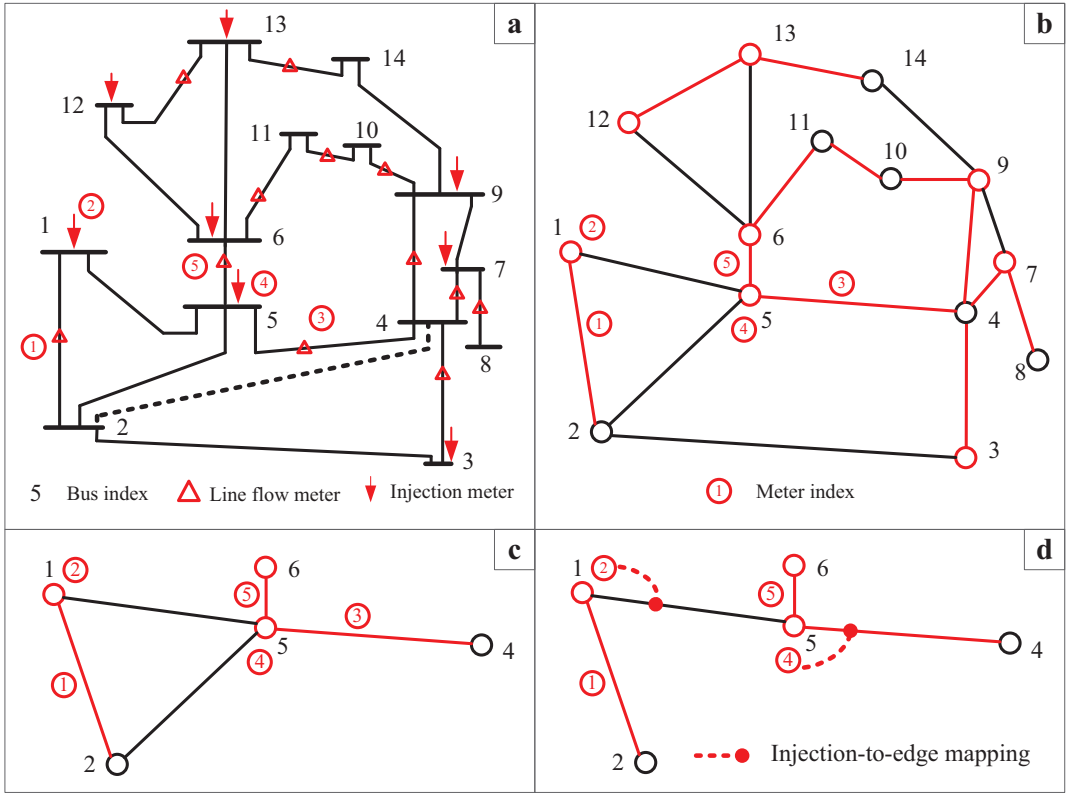}
  \end{center}
  \caption{(a) An example 14-bus power network and measurements; and (b) its graph modeling, where the red vertices (edges) denote the buses (transmission lines) that have injection (flow) meters installed; (c) an example measured subnetwork; and (d) an edge-measured Steiner three embedded in the subnetwork.}
  \label{71}
\end{figure}

As shown in Fig.~\ref{71}(a), a power network consists of a number of buses, loads, power generators, and power transmission lines that interconnect them.\footnote{The topology of the power network in Fig.~\ref{71} is adapted from the IEEE $14$-bus test case system (available online at https://www.ee.washington.edu/research/pstca/, Sept. 2016.)} One important parameter representing the operating state of the power system is the phasor of each bus, including its voltage phase angle and voltage magnitude. In practice, the voltage magnitudes can often be directly measured, while the values of phase angles need to be obtained from state estimation \cite{2004:Abur}. Conventionally, in the linearized DC measurement model, the estimate of the phase angles is obtained from the active power measurements, i.e., the active power flows along the power lines (e.g., meter $1$) and the active power injections at the buses (e.g., meter $2$). In recent years, phasor measurement unit (PMU) has emerged as an advanced metering technology that can provide direct real-time voltage phasor measurement with high accuracy and reliability in addition to the conventional meters. In practice, due to high PMU installation cost and the legacy power system in operation, state estimation is often obtained from a mixture of PMU and power flow measurements.

For a power network with $n+1$ buses, we regard one of them as the reference bus, denoted by $R$, and estimate the phase angles of the rest $n$ buses (state variables) from $m$ meter measurements, denoted by $\boldsymbol{\theta}=\left(\theta_1,\theta_2,..,\theta_n\right)'$ and $\mathbf{z}=\left(z_1,z_2,..,z_m\right)'$, respectively. Besides, we denote the set of $n$ unknown buses as $\mathcal{S}$, the set of all the buses $\mathcal{V}\triangleq R \cup \mathcal{S}$, the set of transmission lines as $\mathcal{E}$, and the set of $m$ measurements as $\mathcal{M}$.

As shown in Fig.~\ref{71}(b), a power network can also be described in an undirected graph, where vertices and edges represent buses and transmission lines, respectively. Without loss of generality, we regard bus $1$ as the reference throughout this article. Loosely speaking, a flow meter reflects the difference between two state variables; an injection meter reflects the sum of differences of a state variable with respect to the subset of state variables in one-hop distance; a PMU meter reflects the difference of a state variable with respect to the reference bus. For the convenience of exposition, we consider in this article only conventional power flow measurements. In fact, a PMU measurement can be equivalently converted to a flow measurement in security analysis, which is discussed in \cite{2014:Bi}.

Given a subset of meter measurements $\mathcal{\bar{M}}\subseteq \mathcal{M}$, we can find correspondingly a subnetwork (and thus a subgraph) measured by $\mathcal{\bar{M}}$, denoted by $G\left(\mathcal{\bar{M}}\right) = \left(\mathcal{\bar{V}}, \mathcal{\bar{E}} \right)$. That is, a flow meter measures the transmission line where it is installed and the two buses in both ends; an injection meter measures the bus it is installed, the transmission lines connected to the bus, and all the buses on the other end of the transmission lines. In Fig.~\ref{71}(c), for instance, the subgraph measured by $\mathcal{\bar{M}} = \left\{r_1,r_2,r_3,r_4,r_5\right\}$ is $\mathcal{\bar{V}} = \left\{v_1,v_2,v_4,v_5,v_6\right\}$ and $\mathcal{\bar{E}} = \left\{e_{12},e_{15},e_{25},e_{45},e_{56}\right\}$, where $r_2$ and $r_4$ are injection meters and $e_{ij}$ denotes the edge connecting vertex $i$ and $j$. For a normal power network, the measured full graph $G\left(\mathcal{M}\right)$ includes all the vertices $\mathcal{V}$ to estimate all the state variables, but not necessarily all the transmission lines. For instance, we can see that the transmission line between bus $2$ and $4$ is not measured by any meter, and thus is not present in the graph model in Fig.~\ref{71}(b).

\section{Graphical Characterization of State Estimation Protection}\label{sec:scalable_structure}

\subsection{State Estimation Problem}
The state estimation problem is to derive the unique estimation of $\boldsymbol{\theta}$ from the measurements $\mathbf{z}$, which are related by
\begin{equation}
\label{1}
\mathbf{z} = \mathbf{H}\boldsymbol{\theta} + \mathbf{e}.
\end{equation}
Here, $\mathbf{H}$ denotes the measurement Jacobian matrix and $\mathbf{e}$ denotes independent measurement noise with zero mean. The exact value of $\mathbf{H}$ is related to the physical aspects of the power network, e.g., the network topology, the placement of meters, and the transmission line impedance \cite{2004:Abur}. In particular, we consider in this article a well-functioning power network that a unique estimate $\boldsymbol{\hat{\theta}}$ of the unknown variables can be obtained from the received measurements. This requires sufficient number of meters to be placed in proper locations such that $\mathbf{H}$ is full column rank, i.e., $rank\left(\mathbf{H}\right)=n$. At least $n$ meters are needed to derive a unique state estimation. Meanwhile, the other $m-n$ measurements provide the redundancy to improve the resistance against random errors. Detailed meter placement methods can be found in \cite{1980:Krumpholz}. Let $\boldsymbol{\hat{\theta}}$ denote the maximum likelihood estimation of $\boldsymbol{\theta}$ \cite{2004:Abur}. The current power systems use BDD mechanism to remove the bad data assuming that the errors are random and unstructured. It calculates the residual $\mathbf{r}=\mathbf{z}-\mathbf{H\boldsymbol{\hat{\theta}}}$ and compares its $l_2$-norm with a prescribed threshold $\tau$. A measurement $\mathbf{z}$ is identified as a bad data measurement if $r=||\mathbf{z}-\mathbf{H\boldsymbol{\hat{\theta}}}||_2> \tau$, or otherwise a normal measurement.

\subsection{Data Injection Attack}
A data injection attack compromises the normal measurements through either physical access or remote cyber control, resulting in fabricated measurements $\mathbf{\tilde{z}} = \mathbf{z}+\mathbf{a}$, where $\mathbf{a}$ denotes the injected data. It can be easily shown that an injection attack structured as $\mathbf{a} = \mathbf{H}\mathbf{c}$, where $\mathbf{c}$ is an arbitrary vector, will produce the same BDD residual as the normal measurement $\mathbf{z}$, thus can introduce a bias $\mathbf{c}$ to the state estimate without being recognized as a malicious attack \cite{2009:Liu}. This kind of attack is commonly referred to as \emph{undetectable attack}. In general, such an attack requires high level of coordination to compromise multiple measurements simultaneously. In some cases, however, the adversary can exploit the special structure of $\mathbf{H}$ to achieve the attacking objective by compromising only a small number of measurements. In fact, we will show later how to use graphical methods to exploit the opportunity of undetectable attack with the minimum number of meter measurements to compromise.

\subsection{Power Network Observability}
State estimation protection is closely related to the concept of power \emph{network observability}. The conventional power network observability analysis studies whether a unique estimate of all unknown state variables can be determined from the measurements \cite{2004:Abur}. Notice that the observability of a network is related to the network topology and the placement of meter measurements, rather than the value of received measurements in real-time. Out of the $m$ total meters, a set of $n$ meter measurements is referred to as a \emph{basic measurement set} if the estimation of $n$ unknown state variables can be uniquely derived from them. It is proved that the presence of any data injection attack can be detected if we can make sure that the measurements taken from at least one basic measurement set are trustworthy, i.e., the meters are well-protected \cite{2010:Bobba}. Intuitively, this is because the estimation obtained from a basic measurement set can be used to validate the result derived from all the meter measurements.

In a large-size power network with several hundred of state variables, it could be infeasible to perform security upgrade to protect $n$ basic measurements under limited budget. Even if sufficient budget is given, protecting the $n$ basic measurements in a random sequence may still open to attackers the possibility to compromise a large number of state variables during the lengthy security installation period. In both cases, it is valuable to devise a method that gives priority to defending a subset of state variables that serve our best interests at the current stage, and opens to the possibility of expanding the set of protected state variables in the future.

In light of this, \cite{2014:Bi} generalizes the concept of power network observability to \emph{subnetwork observability}. Specifically, a subnetwork $G\left(\mathcal{\bar{M}}\right) = \left(\mathcal{\bar{V}}, \mathcal{\bar{E}} \right)$ is referred to as observable if a unique estimation of $\mathcal{\bar{V}}$ can be derived from $\mathcal{\bar{M}}$. Then, protecting the measurements in $\mathcal{\bar{M}}$ can ensure that any data injection attack can be detected as long as it attempts to compromise any member in $\mathcal{\bar{V}}$. The observability of $G\left(\mathcal{\bar{M}}\right)$ can be easily determined with simple matrix calculation. Accordingly, to defend a set of state variables, denoted by $\mathcal{D}$, the problem becomes finding an optimal observable subnetwork $G\left(\mathcal{\bar{M}}\right)= \left(\mathcal{\bar{V}}, \mathcal{\bar{E}} \right)$, either with the minimum number of vertices or the minimum cost to secure the meters in $\mathcal{\bar{M}}$, that satisfies $\mathcal{D} \subseteq \mathcal{\bar{V}}$. An intuitive solution is to enumerate all possible vertices in $\mathcal{S}\setminus \mathcal{D}$ to check if an observable subnetwork can be constructed together with $\mathcal{D}$. This enumeration method, however, is combinatorial in nature, and indeed the problem to find the optimal subnetwork is proved to be NP-Hard \cite{2014:Bi}.

\subsection{Graphical Characterization of Observability}
Alternatively, the network observability has an intuitive characterization using graphs. Specifically, a subnetwork $G\left(\mathcal{\bar{M}}\right) = \left(\mathcal{\bar{V}}, \mathcal{\bar{E}} \right)$ is observable if and only if an \emph{edge-measured Steiner tree} (EMST) \cite{2014:Bi}, denoted by $T=\left(\mathcal{\bar{V}},\mathcal{\hat{E}}\right)$, can be constructed from the subnetwork and satisfies the following conditions:
\begin{enumerate}
  \item the reference vertex $R$ is contained in the tree, i.e., $R\in \mathcal{\bar{V}}$;
  \item each edge $e\in \mathcal{\hat{E}}$ is mapped to a flow meter or an injection meter $p \in \mathcal{\bar{M}}$ that measures it;
  \item different edges are mapped to different meters in $\mathcal{\bar{M}}$.
\end{enumerate}
Intuitively, this requires to find a tree that connects all the vertices in the subgraph to the reference vertex, where each edge is mapped to a meter that takes its measurement. For instance, an EMST and the measurement-to-edge mappings are shown in Fig.~\ref{71}(d) for the observable subnetwork in Fig.~\ref{71}(c). Such a tree is named a Steiner tree because in general only a subset of vertices is included in the tree. A special case is $\mathcal{\bar{V}}=\mathcal{V}$, where the Steiner tree becomes a \emph{spanning tree} that includes all the vertices in the network \cite{1980:Krumpholz}. Thanks to the graphical structure of an observable subnetwork, we introduce in the following section some efficient graphical algorithms for security analysis in power system.

\section{Graph Algorithms for Power System Security Analysis}\label{sec:algorithm}

\begin{figure}
\centering
  \begin{center}
    \includegraphics[width=0.6\textwidth]{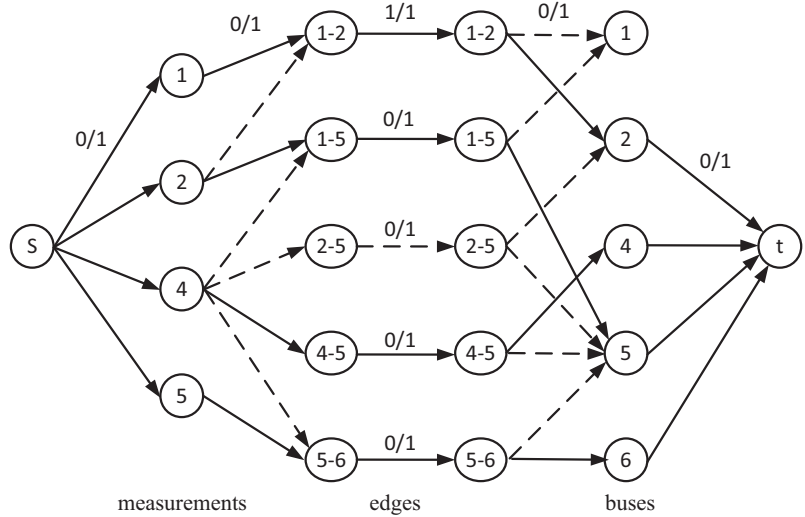}
  \end{center}
  \caption{Illustration of maximum flow method for constructing an EMST from an observable subnetwork. The solid lines denote saturating edges while the dashed lines denote unused edges.}
  \label{72}
\end{figure}

\subsection{Maximum-flow Matching Algorithm}
The graphical characterization establishes the equivalence between the subnetwork observability and the existence of an embedded EMST. A natural question is how to construct such an EMST from an observable subnetwork $G\left(\mathcal{\bar{M}}\right) = \left(\mathcal{\bar{V}}, \mathcal{\bar{E}} \right)$, which is very useful in visualizing the network observability to enable efficient tree-based algorithms. As finding a set of meters $\mathcal{\hat{M}} \subseteq \mathcal{\bar{M}}$ to derive a unique estimation of $\mathcal{\bar{V}}$ is easily achievable through Gauss-Jordan matrix elimination, the question lies in how to find the mappings between $\mathcal{\hat{M}}$ and the edges $\mathcal{\bar{E}}$ to satisfy the EMST definitions. Interestingly, the EMST construction problem can be solved in polynomial time using a \emph{maximum-flow} method \cite{1986:Barglela}.

We use an observable subnetwork in Fig. \ref{71}(c) as an example to illustrate the method to obtain an EMST. As shown in Fig.~\ref{71}(d), we have $\mathcal{\bar{V}}=\left\{v_1,v_2,v_4,v_5,v_6\right\}$, $\mathcal{\hat{M}}=\left\{r_1,r_2,r_{4},r_{5}\right\}$ and the set of edges measured by $\mathcal{\hat{M}}$ is $\mathcal{\bar{E}}=\left\{e_{12},e_{15},e_{25},e_{45},e_{56}\right\}$. Then, a directed graph is constructed in Fig. $\ref{72}$, where $v_1$ is chosen as the root to construct the Steiner tree. We select in advance an edge connected to the root, say $e_{12}$, in the final tree solution. This is achieved by setting both the lower and upper capacity bounds of the edge to be $1$. The other edges' lower and upper capacity bounds are set to be $0$ and $1$, respectively. Then, a maximum flow is calculated from the source ($s$) to the terminal ($t$), which is achievable in polynomial time using, e.g., Ford-Fulkerson Algorithm \cite{1986:Barglela}. If the problem is feasible, i.e., the flow solution is $1$ in edge $e_{12}$, we obtain a measurement-to-edge mapping by observing the saturating flows in the graph. Otherwise, the actual EMST solution does not include $e_{12}$ (i.e., the initial guess is wrong), thus we select another edge connected to the root and recalculate the maximum flow problem. Since the subnetwork is observable, the existence of a solution is guaranteed. In the above example, the final measurement-to-edge mapping is $\left\{r_1,r_2,r_{4},r_{5}\right\}\leftrightarrow \left\{e_{12},e_{15},e_{45},e_{56}\right\}$, while edge $e_{25}$ is not used. Then, the edges obtained by the maximum flow calculation will form a tree that spans all vertices in $\bar{V}$ as shown in Fig.~\ref{71}(d).

\subsection{Commodity Flow Maximization Algorithm}
Although finding a minimum EMST that includes a set of vertices $\mathcal{D}$ is NP-Hard, a \emph{commodity flow formulation} that exploits the tree structure of EMST can largely reduce the complexity compared to some enumeration based methods, e.g., from several months to a couple of minutes in a medium-size network. Intuitively, this is because the graph-based formulation can significantly reduce the search space of candidate solutions and enable effective off-the-shelf graph/optimization algorithms.

\begin{figure}
\centering
  \begin{center}
    \includegraphics[width=0.75\textwidth]{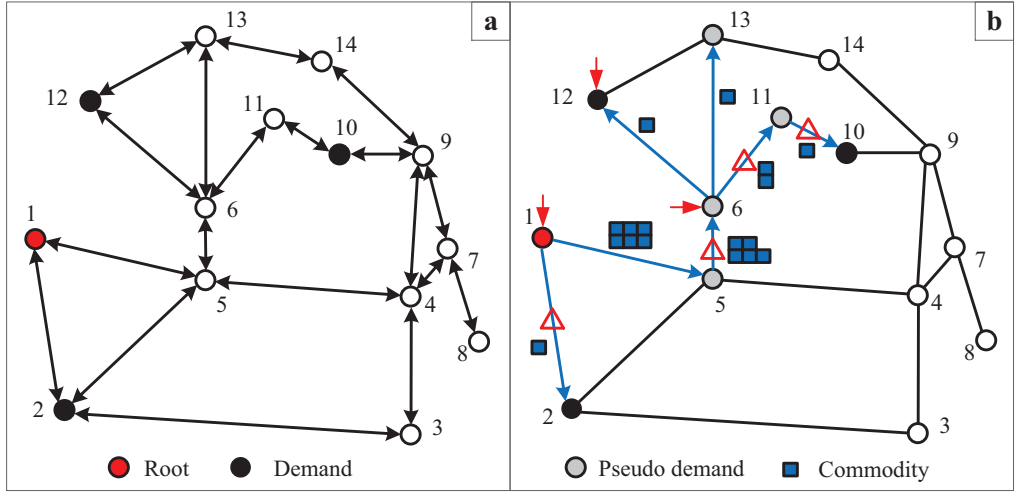}
  \end{center}
  \caption{Illustration of commodity flow maximization method for solving the optimal EMST problem. a) the Steiner arborescence constructed from the graph model in Fig.~\ref{71}(b); and the maximum commodity flow solution when the arc weight is the same for all the arcs.}
  \label{73}
\end{figure}

Consider a digraph $\overrightarrow{G}=\left(\mathcal{V},\mathcal{A}\right)$ constructed by replacing each edge in the measured full graph $\bar{G}\left(\mathcal{M}\right)=\left(\mathcal{V},\mathcal{E}\right)$ with two arcs in opposite directions. We set the reference bus as the root and allocate one unit of demand to each vertex in $\mathcal{D}$. As shown in Fig.~\ref{73}, commodities are sent from the root to the vertices in $\mathcal{D}$ through some arcs. Notice that the choice of the vertices $\mathcal{D}$ in Fig.~$\ref{73}$ is only for the simplicity of illustration, where an arbitrary subset of vertices $\mathcal{D}\subseteq \mathcal{S}$ can be selected. Then, the vertices in $\mathcal{D}$ are connected to $R$ via the used arcs if and only if all the demands are satisfied. When we require using the minimum number of arcs to deliver the commodity, the used arcs will form a directed tree, referred to as a \emph{Steiner arborescence}. Evidently, the solution to the minimum EMST problem can be obtained if we neglect the orientations of the arcs in the obtained Steiner arborescence. To satisfy the conditions of a feasible EMST, we need to make sure that any selected arc is mapped to a meter that measures it. In particular, if an arc is mapped to an injection meter, all the vertices measured by the injection meter must also be included in the arborescence, as if a pseudo demand is allocated at these vertices. Then, the problem is to satisfy both the actual and pseudo demand using minimum number of arcs.

Based on the commodity flow model, a mixed integer linear programming (MILP) formulation is proposed in \cite{2014:Bi}, and extended to arcs of different weights (different costs are needed to secure the meters) in \cite{2014:Bi1}, which can be solved with many off-the-shelf integer optimization tools, such as Gurobi and CPLEX. Accordingly, we can use the mappings from the arcs in the optimal EMST to the optimal set of meter measurements that defends the state variables in $\mathcal{D}$.

\subsection{Tree Pruning Algorithm}
Due to the NP-Hardness of finding an optimal EMST, the commodity flow based method can still result in high computational complexity in a large-size power network consisting of hundreds of buses. A polynomial-time suboptimal algorithm using the idea of tree pruning is considered in \cite{2014:Bi}. Starting from the full measured graph, the key idea is to iteratively construct an EMST from the subnetwork and prune away redundant vertices not in $\mathcal{D}$, while keeping the remaining subnetwork formed by the residual vertices observable until a shortest possible EMST is obtained. Specifically, the \emph{tree traversal} algorithm can be applied to determine both the sequence and the subset of vertices to be pruned in each iteration.

\begin{figure}
\centering
  \begin{center}
    \includegraphics[width=0.75\textwidth]{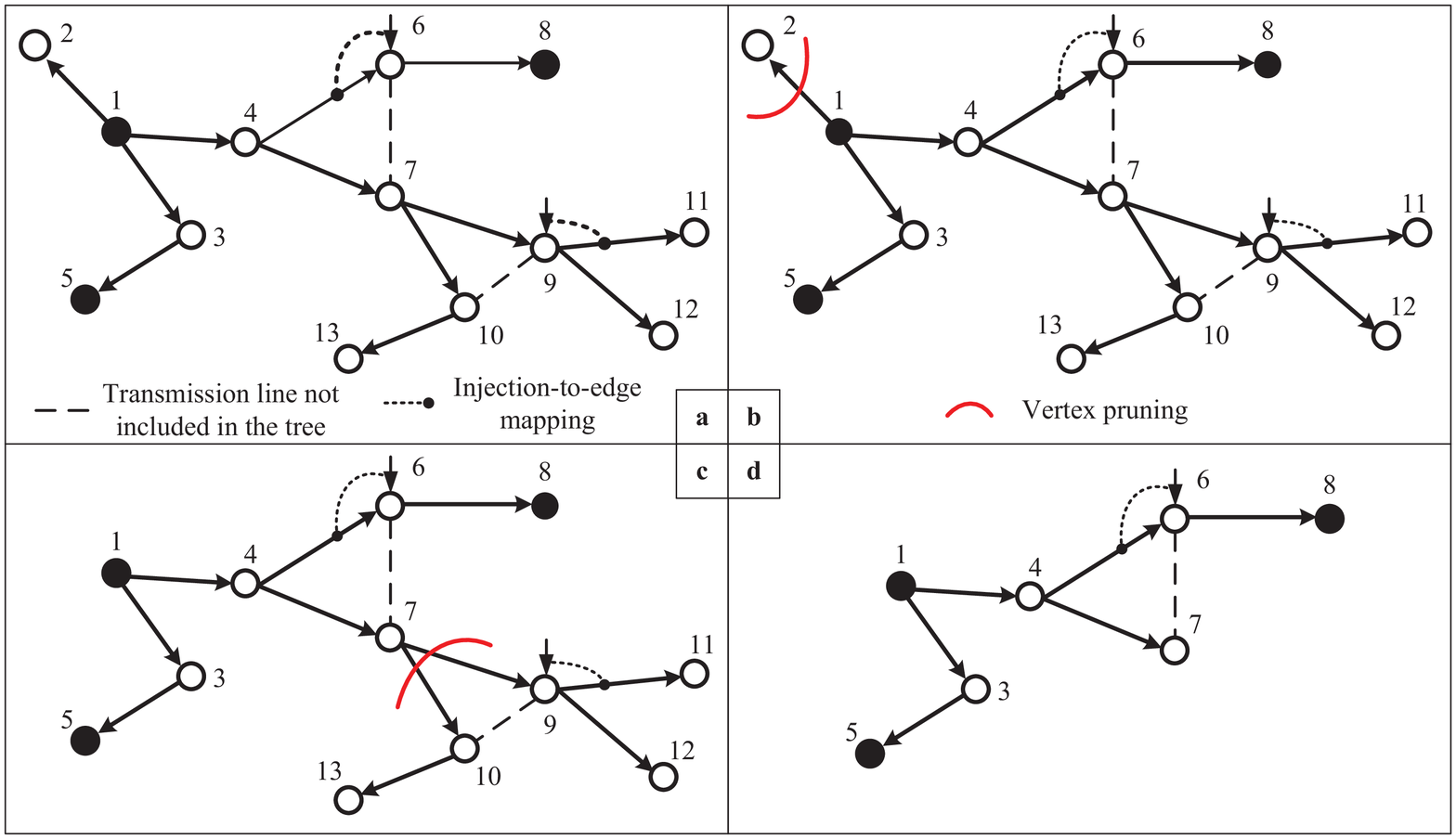}
  \end{center}
  \caption{Illustration of the tree pruning algorithm. The shaded vertices are terminals to be included in the EMST.}
  \label{74}
\end{figure}

In Fig. \ref{74}, we present an example to illustrate the pruning operation, where a feasible tree containing $12$ vertices is presented in Fig. \ref{74}(a). Vertex $5$ and $8$ are the \emph{terminal} vertices to be included in the EMST solution. As shown in Fig. \ref{74}(b), starting from the root $v_1$, among the three child vertices of $v_1$, only $v_2$ can be pruned, since the descendent vertices of either $v_3$ or $v_4$ contain terminal vertex. After pruning $v_2$, we proceed to check $v_3$ to see if its child vertex $v_5$ can be pruned, which, however, is not feasible because $v_5$ is a terminal. Then, we check $v_4$, where neither of its child vertices $v_6$ and $v_7$ can be pruned separately or together. On one hand, this is because $v_6$ contains terminal as its descendent vertices. On the other hand, the removal of $v_7$ does not remove the edge $e_{46}$, which is mapped to the injection meter at $v_6$ that measures $v_7$, thus resulting an unobservable residual subnetwork. For $v_7$, however, all of its descendent vertices can be pruned as in Fig. \ref{74}(c). Up to now, we have finished the first round of pruning and obtained a residual tree in Fig. \ref{74}(d). Then, we use the remaining vertices $\left\{v_1,v_3,v_4,v_5,v_6,v_7,v_8\right\}$ to generate new EMSTs using the maximum-flow matching algorithm and repeat the pruning operations iteratively until no vertex can be further pruned.

It is shown in \cite{2014:Bi} that the tree pruning heuristic (TPH) can achieve comparable performance with the optimal solution obtained from the commodity flow MILP formulation especially in a large-size network. Meanwhile, it induces much lower complexity. For instance, using a regular computer with Intel Core2 Duo 3.00-GHz CPU and 4 GB of memory, the average computation time needed to solve for $|\mathcal{D}|=4$ buses out of a $14$-bus network is $\left\{0.04,0.2,0.02\right\}$ seconds for the arithmetic-based enumeration, the introduced MILP formulation and the TPH methods, respectively \cite{2014:Bi}. However, the computation time of arithmetic-based enumeration grows dramatically to around $90$ years to solve for $|\mathcal{D}|=4$ in a $57$-bus network, which is computationally infeasible in practice. This, however, takes the MILP and the TPH methods only $3.7$ and $0.12$ seconds, respectively. As we further increase the size to a $118$-bus network, the computation time of the TPH method increase almost linearly to $0.49$ second, while the optimal MILP formulation increases quickly to around $5$ minutes. In this sense, the TPH method can efficiently solve a problem in very large networks of several hundred of buses within a couple of seconds, which may take the MILP method many days or even months to complete.

\subsection{Minimum S-T Cut Algorithm}
\begin{figure}
\centering
  \begin{center}
    \includegraphics[width=0.75\textwidth]{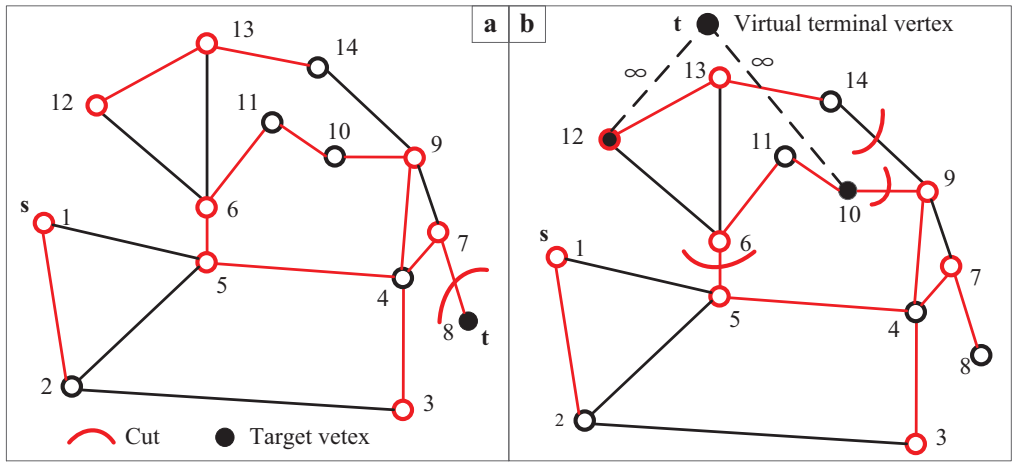}
  \end{center}
  \caption{Illustration of minimum S-T cut algorithm to exploit cyber vulnerability. The edge weight is $1$ for each edge unless otherwise stated. Specifically, figure (a) shows the cut to attack a single bus $8$; and (b) shows the cut to compromise two buses $10$ and $12$. In (b), a virtual terminal is added to connect the target vertices, while the edge weights between them are set as infinity to avoid a cut on any of them.}
  \label{76}
\end{figure}

An adversary can also apply graphical methods to exploit the opportunity to launch malicious attacks. A widely used algorithm is \emph{minimum S-T cut} method, which calculates the minimum sum weights of edges, whose removal would separate a source vertex from a terminal vertex in a weighted graph \cite{2012:Sou}. Intuitively, an adversary that intends to compromise a state variable will need to separate the corresponding vertex (the terminal) from the reference vertex (the source) in the graph by forming a cut on the edges. Then, the adversary needs to compromise all the meters that measure the edges in the cut. For instance, in Fig.~\ref{76}(a), the cut on $e_{78}$ to attack bus $8$ requires the adversary to compromise the flow meter on edge $e_{78}$ and the injection meter on bus $7$. The weight of each edge in the calculation of the minimum S-T cut problem can be set as the monetary cost to compromise the meters that measure it.

Similar minimum cut methods can also be applied to compromise a set of state variables \cite{2014:Bi} (see Fig.~\ref{76}(b)); to find the smallest number of meters that the adversary can control to perform an unobservable attack \cite{2011:Kosut}; to identify the most vulnerable measurements to inject false data \cite{2012:Sou}; and to exploit the opportunity of data injection attack when some meters are secured or the network topology is only partially known \cite{2012:Rahman,2014:Bi1}. As minimum S-T cut can be efficiently calculated in polynomial time, an adversary is able to quickly identify potential network security vulnerability.

\section{Future Research Directions}\label{sec:future}

\subsection{Application Oriented Security Analysis}
Essentially, the power system state estimation is used for controlling specific applications, such as generation/load power control and electricity price calculation. It is therefore of practical value to perform application-oriented security analysis in higher application layer. Existing studies have shown that data injection attacks that cause blackout and electricity price manipulation have apparent graphical patterns \cite{2013:Bi,2011:Yuan}. It is therefore interesting to exploit the underlying graphical structures in the attacks to compromise power applications, such as load prediction, unit commitment, and frequency control. On the other hand, it is also useful to use graphical methods to strategically deploy security countermeasures, e.g., to prevent collaborating attacks that compromise the electricity market.

\subsection{Meter Measurement Placement Optimization}
As we are now transforming the legacy power system to the future smart grid, a large amount of electricity infrastructure are to be built within the near future, with a mixture of conventional and new metering/communication facilities. Many existing security vulnerabilities often comes from the legacy meter measurements placement, which hardly considers the threat of potential collaborating attacks. Graphical methods can be useful to optimize the placement of the meter measurement. By leveraging the graphical properties of network observability, we have the potential to achieve both high state estimation accuracy and high resistance to potential data attacks in relatively low meter placement cost.

\subsection{Hybrid Graphical and Data-driven Approaches}
Graph-based security analysis is an offline ``hardware" approach, where physical protections are performed to ensure the measurements collected from a subset of meters are trustworthy (free from injection attacks). Data-driven attack detection, on the other hand, is an online ``software" approach that leverages the statistical features of the measurements/state variables to identify potential abnormal measurements collected from the rest unsecured meters. In particular, graph-based method is independent of real-time measurements and does not alter the state estimation algorithm in EMS/SCADA. Therefore, it can be potentially combined with data-driven detection to further improve system security. For instance, the trust-worthy measurements, and so the subset of trust-worthy state estimates derived from them, can be used as side information to improve the detection accuracy of data-driven statistical detections. In general, the graph-based protection and data-driven method should be jointly designed.

\subsection{Security Analysis in AC Model}
Graph algorithms are commonly used to solve linear integer programming problems, their effectiveness and efficiency to solve security problem in linear DC power system is unsurprising. In many application scenarios, however, AC power model, where both voltage amplitude and phase are the state variables, is more preferable than the DC model, e.g., security constrained optimal power flow calculation. Some studies have shown that data injection attack to compromise AC state estimation is much more complicated than that in DC model \cite{2012:Hug}. On the other hand, the observability of AC state estimation can no longer be characterized as a simple Steiner tree structure as in DC model. However, the network observability may still contain tree-like structures to be identified for defending potential attacks against AC state estimation.

\section{Conclusions}\label{sec:conclusion}
In this article, we have provided a graphical framework for performing security analysis in power system state estimation. From both system operator's and adversary's perspectives, we have introduced several effective graph-based algorithms to solve security problems in state estimation. Compared to the commonly used arithmetic-based security analysis, graph-based analysis helps visualize some complex problem structures, which can lead to efficient optimal or reduced-complexity suboptimal graph-based algorithms. As the future smart power grid will integrate a large number of ICT facilities, cyber security is of paramount importance to guarantee the system consistently operating in a secure and efficient state. Graph-based methods are expected to be a set of powerful tools in solving complex cyber security problems in future smart grid.


\begin{thebibliography}{1}
\small

\bibitem{2004:Abur}
A.~Abur and A.~G.~Exp\'{o}sito, ``Power system state estimation: theory and implementation", New York: Marcel Dekker, 2004.

\bibitem{2009:Liu}
Y.~Liu, P.~Ning, and M.~Reiter, ``False data injection attacks against state estimation in electric power grids," in \emph{Proc. ACM CCS}, Chicago, USA, Oct. 2009.

\bibitem{2013:Bi}
S.~Bi and Y.~J.~Zhang, ``False data injection attack to control real-time price in electricity market," in \emph{Proc. IEEE Globecom}, Atlanta, USA, Dec. 2013.

\bibitem{2011:Kosut}
O.~Kosut, L.~Jia, R.~J.~Thomas, and L.~Tong, ``Malicious data attack on the smart grid", \emph{IEEE Trans. Smart Grid}, vol.~2, no.~4, pp.~645-658, Dec.~2011.

\bibitem{2011:Yuan}
Y.~Yuan, Z.~Li, and K.~Ren, ``Modeling load redistribution attacks in power systems," \emph{IEEE Trans. Smart Grid}, vol.~2, no.~2, pp.~382-390, Jun. 2011.

\bibitem{2013:Huang}
Y.~Huang, M.~Esmalifalak, H.~Nguyen, R.~Zheng, Z.~Han, H.~Li, and L.~Song, ``Bad data injection in smart grid: attack and defense mechanisms," \emph{IEEE Commun. Mag.}, vol.~51, no.~1, pp.~27-33, Jan. 2013.

\bibitem{2010:Bobba}
R.~Bobba, K.~M.~Rogers, Q.~Wang, H.~Khurana, K.~Nahrstedt and T.~Overbye, ``Detecting false data injection attacks on DC state estimation," in \emph{Proc. CPSWEEK}, Apr.~2010.

\bibitem{2011:Bi}
S.~Bi and Y.~J.~Zhang, ``Defending mechanisms against false-data injection attacks in the power system state estimation," in \emph{Proc. IEEE Globecom}, Houston, USA, Dec. 2011.


\bibitem{2014:Bi}
S.~Bi and Y.~J.~Zhang, ``Graphical methods for defense against false-data injection attacks on the power system state estimation," \emph{IEEE Trans. Smart Grid}, vol.~5, no.~3, pp. 1216-1227, May 2014.

\bibitem{2012:Sou}
K.~C.~Sou, H.~Sandberg, and K.~H.~Johansson, ``Computing critical k-tuples in power networks", \emph{IEEE Trans. Power Syst.}, vol.~27, no.~3, pp.~1511-1520, Aug. 2012.

\bibitem{2012:Rahman}
M.~A.~Rahman and H.~Mohsenian-Rad, ``False data injection attacks with incomplete information against smart power grids," in \emph{Proc. IEEE Globecom}, Anaheim, USA, Dec.~2012.

\bibitem{2014:Bi1}
S.~Bi and Y.~J.~Zhang, ``Using covert topological information for defense against malicious attacks on DC state estimation," \emph{IEEE J. Sel. Areas Commun.}, vol.~32, no.~7, pp.~1-15, Jul. 2014.

\bibitem{1980:Krumpholz}
G.~R.~Krumpholz, K.~A.~Clements, and P.~W.~Davis, ``Power system observability: a practical algorithm using network topology," \emph{IEEE Trans. Power App. Syst.}, vol.~PAS-99, no.~4, pp.~1534-1542, Jul.~1980.

\bibitem{1986:Barglela}
A.~Barglela, M.~R.~Irving, and M.~J.~H.~Sterling, ``Observability determination in power system state estimation using a network flow technique", \emph{IEEE Trans. Power Syst.}, Vol.~PWRS-1, No.~2, May 1986.

\bibitem{2012:Hug}
G.~Hug and J.~A.~Giampapa, ``Vulnerability assessment of AC state estimation with respect to false data injection cyber-attacks," \emph{IEEE Trans. Smart Grid}, vol.~3, no.~3, pp.~1362-1370, Sept.~2012.

\end{thebibliography}
\end{document}